# CRYSTAL UNDULATOR AS A NOVEL COMPACT SOURCE OF RADIATION

S. Bellucci, S. Bini, G. Giannini, *INFN-LNF Frascati, Italy*; V.M. Biryukov, G.I. Britvich, Yu.A. Chesnokov, V.I. Kotov, V.A. Maisheev, V.A. Pikalov, *IHEP Protvino, Russia*; V. Guidi, C. Malagù, G. Martinelli, M. Stefancich, D. Vincenzi, *Univ. Ferrara, INFM-INFN Italy*; Yu.M. Ivanov, A.A. Petrunin, V.V. Skorobogatov, *PNPI, St. Petersburg, Russia*; F. Tombolini; *Univ. Roma Tre, Italy*

*Abstract*

A crystalline undulator (CU) with periodically deformed crystallographic planes is capable of deflecting charged particles with the same strength as an equivalent magnetic field of 1000 T and could provide quite a short period L in the sub-millimeter range. We present an idea for creation of a CU and report its first realization. One face of a silicon crystal was given periodic micro-scratches (grooves), with a period of 1 mm, by means of a diamond blade. The X-ray tests of the crystal deformation have shown that a sinusoidal-like shape of crystalline planes goes through the bulk of the crystal. This opens up the possibility for experiments with high-energy particles channeled in CU, a novel compact source of radiation. The first experiment on photon emission in CU has been started at LNF with 800 MeV positrons aiming to produce 50 keV undulator photons.

## INTRODUCTION

The energy of a photon, $E$, emitted in an undulator is in proportion to the square of the particle Lorentz factor $\gamma$ and in inverse proportion to the undulator period L: $E = h\gamma^2 c/L$. Typically, at the modern accelerators the period of undulator in the synchrotron light sources is a few centimeters [1].

With a strong worldwide attention to novel sources of radiation, there has been broad interest [2-12] to compact crystalline undulators. A CU with periodically deformed crystallographic planes bends the charged particles as an equivalent electromagnetic field in the order of 1000 T and could provide a period L in sub-millimeter range. This way, a hundred-fold gain in the energy of emitted photons is reached, as compared to a usual undulator.

## PECULIARITIES OF CRYSTALLINE UNDULATOR

Particle trajectories in a deformed crystal are more complicated than in a usual undulator (Fig. 1). Undulator radiation is accompanied by a harder component (channeling radiation). This component is harder than undulator radiation because of the smaller period of oscillations $L_{chr} \sim 3$ micron (for typical energies of a few GeV, where undulators are used). On the other hand it is lower in intensity because the amplitude of oscillations is also small $A_{chr} \sim 1$ Å. At the condition $A_{cu} \gg A_{chr}$ the undulator radiation has much higher spectral density and the background radiation is not essential.

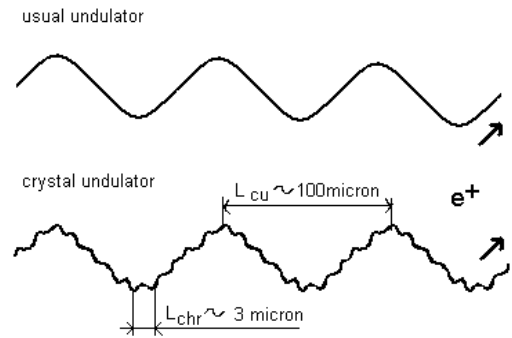

**Fig.1 Peculiarities of a crystalline undulator.**

## METHOD OF CU CREATION

Different ideas have been proposed for creation of CU [2-15], but they are still pending realization. We have recently [2] demonstrated by means of X-rays that microscratches on the crystal surface make sufficient stresses for creation of a CU by making a series of scratches with a period of 1 mm. Now we have optimized this process and were able to produce an undulator with a period in sub-millimeter range and with a good amplitude. A series of undulators was manufactured with the following parameters: length along the beam 1 to 5 mm, thickness across the beam 0.3 to 0.5 mm, 10 periods of oscillation with step from 0.1 to 0.5 mm, amplitude on the order of 50 Å. The undulators were tested with X-rays as described in ref. [2].

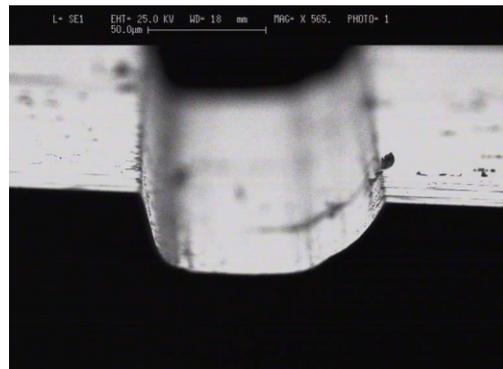

**Fig.2 Example of a microgroove on crystal surface.**

The X-ray (17.4 keV) beam was collimated to 2 mm height and 40 μm width before incidence on the sample surface. The sample could be translated with accuracy of 1 μm and 1" by use of a standard theodolite. A NaI counter with wide-acceptance window detected the diffracted radiation. The count rate of diffracted quanta is maximal under Bragg condition, achieved by rotation of the sample. Fig.3 shows the measured angles as functions of the beam incidence position at the crystal surface. On the same absolute scale, the position of grooves is shown as well. The periodic angular deformations of the crystal planes reach an order of 40-50 μrad. The plane deformation amplitude is in the order of 20 Å as obtained by analysis of the angle-versus-position function of Fig. 3, measured on the opposite (unscratched) face of the crystal. This means that a sinusoidal-like shape of crystalline planes goes through the bulk of the crystal.

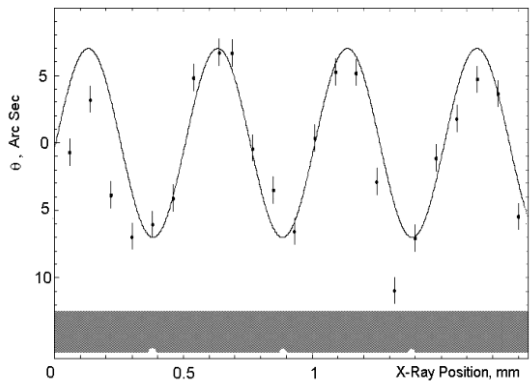

**Fig. 3 X-ray test of one of the undulators.**

## SCHEME OF A PHOTON EMISSION EXPERIMENT

Our collaboration has two appropriate sites for an accelerator experiment on generation of photons in a crystalline undulator. These are LNF with the positron beam energy 500-800 MeV, and IHEP where one can arrange positron beams with energy above 2 GeV. In order to improve background condition, one needs to place the CU into vacuum. A goniometer provides an angling of crystal within ±20 mrad with a step size of about 0.050 mrad. Right after the vacuum box, a cleaning magnet is positioned. The vacuum system is ended by a tube as long as 3 m, 200 mm in diameter, which has at the end a Mylar window 0.1-0.2 mm thick.

As a detector of photons, we use a crystal of NaI (Tl) with a diameter of 1 cm ×10 cm. To calibrate the γ-detector, we use radioactive sources $^{241}$Am with $E_\gamma$=59 keV, and $^{60}$Co with $E_\gamma$=1.15 MeV. Aim of the experiment is the observation of undulator photons emitted with expected energy in a CU, measurement of its spectrum, experimental comparison to the case of a usual straight crystal, and finally demonstration that the crystalline undulator works. First experimental results are expected soon, subject to the schedule of our accelerators.

## EXPECTED PHOTON SPECTRUM

The calculations of radiation intensity were carried out for the (011) plane of the silicon single crystal. It is obvious that only positrons under channeling conditions can emit the low frequency radiation in the periodic above-considered structure. Thus, the channeling radiation will take place for similar crystal structures. In general, radiation is characterized by dimensionless parameter [14] $\rho = 2 \gamma^2 <v_p^2>/c^2$, where $<v_p^2>$ is the mean squared transversal velocity of the particle. For CU, $<v_p^2> \approx <v_u^2> + <v_c^2>$ if the curvature of trajectory is not large. Here $<v_u^2>$ and $<v_c^2>$ are mean squared transversal velocities for undulator and channeling motion, correspondingly.

When $\rho \leq 1$, the total radiation spectrum is a simple additive combination of the strictly undulator and channeling ones. The total spectrum will be the sum of contributions from mainly two basic frequencies of the both processes. When $\rho >> 1$ one can expect that the spectrum will be similar to photon spectrum of synchrotron radiation.

For ρ slightly more than 1, the radiation spectrum is complicated and consists of some peaks. The last case is more difficult for consideration. Besides, in CU, the dechanneling process and radiation of the above-barrier positrons is expected, and it is necessary to take into account their influence on photon spectrum. For calculations of the photon spectrum from 800 MeV positron beam we chose the period and amplitude of deformation of the silicon single crystal equal to 0.01 cm and 80 Å, respectively; we expect to obtain these values in nearest future.

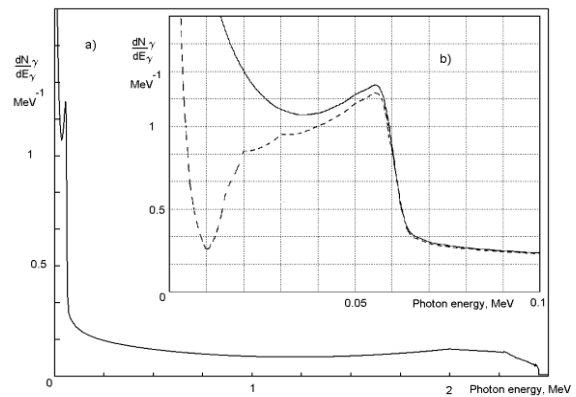

**Fig.4 Expected photon spectrum for 800 MeV positrons in range (0 – 2.5) MeV (a) and (0 – 0.1) MeV (b). The dashed curve is for photon absorption in the body of undulator taken into account. Normalized on one positron incident within channeling angle.**

When ρ is about or below 1, one can calculate the expected photon spectra for LNF experiment.

Fig. 4 shows the calculated spectrum of photons radiated in CU for 800-MeV positrons. The maximum of the distribution corresponds to ≈ 55 keV. Our calculations take into account the following factors:

1) channeling radiation and dechanneling process;
2) finite length of the crystalline undulator, 0.1 cm;
3) radiation of the above-barrier positrons;
4) absorption of gamma-quanta in the undulator bulk (calculated in assumption that absorption in undulator is similar to that in amorphous silicon media [15]).

The channeling radiation is computed in accordance with the paper [16] and these results are in a good agreement with experimental data [17]. Our calculations allow one to make the following conclusions concerning the photon spectrum.

(a) Clear peak of the photon number one can observe in the range 30-60 keV. The sum of the photons within this peak is equal to approximately 20 % of all the photon spectrum (or 0.05 photons per positron). The major part of the photons owe to channeling radiation. These photons are distributed in the wide range (up to 2.5 MeV) and their spectral density (per MeV) is 5 times less than in the range where the undulator photons are.

(b) The influence of the finite length of the single crystal and the dechanneling process are crucial for the density and the form of the undulator photons' spectrum: the density decreased 1.5-2 times and the maximum of the distribution shifted from 61 keV to 55 keV.

(c) The contribution of radiation of the above-barrier positrons is negligible, thereby a sophisticated collimation of the positron beam is not required.

(d) The strong absorption process at energies < 30 KeV allows to obtain more monochromatic undulator photons.

Positrons energies at IHEP accelerator ( beamline 4) are 2–15 GeV. In this case, ρ≈ 1. Our calculations were carried out for the (011) plane of a silicon single crystal. The thickness of the crystal undulator, amplitude and period of the deformations were 0.3 cm, 40 Å and 0.015 cm, respectively. The calculated number of photons in the range (100–600) keV is 0.15 per one positron passing through the crystalline undulator.

At positron beam energies higher than 3 GeV and selected parameters of the crystal undulator the values of ρ > 1 are achieved and can run up to 100 (for energies 10–15 GeV). For a usual undulator this case was solved analytically [14]. However, for CU (where there are two practically independent frequencies of particle motion) finding an analogous solution is an important actual problem. More detailed information concerning the calculations will be published elsewhere.

## CONCLUSIONS

Our studies on the creation and characterisation of the periodically deformed crystalline structures, and calculations of the expected photon spectra allow us to draw the conclusion that crystalline undulator will be able to produce intense X-rays of 10 to 1000 keV. Crystalline undulator would allow to generate photons with the energy on the order of 1 MeV at the synchrotron light sources where one has at the moment only 10 keV, and for this reason crystal undulators have great prospects for application.

## ACKNOWLEDGEMENTS

This work was partially supported by INFN - Gruppo V, as NANO experiment, by INTAS-CERN Grant No. 132-2000 and RFBR Grant No. 01-02-16229, by the "Young Researcher Project" of the University of Ferrara.

## REFERENCES

[1] Beam Line, V. 32, N 1, 2002
[2] S. Bellucci et al., Phys. Rev. Lett. 90 (2003) 034801
[3] V.V. Kaplin, S.V. Plotnikov, and S.A. Vorobiev, Zh. Tekh. Fiz. 50, 1079-1081 (1980).
[4] V.V. Kaplin, S.V. Plotnikov, and S.A. Vorobiev, in: Abstracts of the 10-th Conference on the Application of Charged Particle Beams for Studying the Composition and Properties of Materials, Moscow, 1979, p.28.
[5] V.G. Baryshevsky, I.Ya. Dubovskaya, and A.O. Grubich, Phys. Lett., 77A, 61-64 (1980)
[6] H. Ikezi, Y. Lin, and T. Ohkawa, Phys. Rev., B30, 1567-1568 (1984).
[7] S.A. Bogacz and J.B. Ketersom, J. Appl. Phys. 60, 177-188 (1986).
[8] G.B.Dedkov, Phys.Stat.Sol.(b)184, 535-542 (1994).
[9] A.V. Korol, A.V. Solovev, and W. Greiner, Intern. Journal of Mod. Phys., 8, 49-100 (1999).
[10] U. Mikkelsen and E. Uggerhoj, Nucl. Instr. and Meth., B160, 435 (2000).
[11] R.O. Avakian, et al. NIM B173, 112 (2001)
[12] R.O. Avakian, et al., NIM A 492, 11 (2002).
[13] V.M. Biryukov, Yu.A. Chesnokov, and V.I. Kotov, Crystal Channeling and Its Application at High-Energy Accelerators (Springer: Berlin, 1997)
[14] V.N. Baier, V.M. Katkov, V.M. Strakhovenko, Electromagnetic processes at high energies in oriented single crystals. World Scientific, 1998
[15] J.H. Hubbell, Natl. Stand. Ref. Data ser. NSRDS-NBS 29 (1969)
[16] V.A. Maisheev. Nucl.Instr.Meth. B254, 42, (1996)
[17] J. Bak, et. al. Nucl. Phys.B254;491, (1985).